\documentclass[prb,twocolumn,superscriptaddress]{revtex4-2}
\usepackage{amsmath}
\usepackage{amssymb}
\usepackage{amsthm}
\usepackage{amsfonts}
\usepackage[version=4]{mhchem}
\usepackage{listings}
\usepackage{enumerate}
\usepackage{latexsym}
\usepackage{psfrag}
\usepackage{bm}
\usepackage[all]{xy}
\usepackage{subfigure}
\usepackage[pdftex,colorlinks]{hyperref}
\usepackage{color}
\usepackage{mathtools}
\usepackage{times}
\usepackage{array}
\usepackage{placeins}

\begin{document}

\title{Dzyaloshinskii-Moriya anisotropy effect on field-induced magnon condensation in kagome antiferromagnet \ce{$\alpha$-Cu3Mg(OH)6Br2} }
\author{Ying Fu}
\thanks{These authors contributed equally.}
\affiliation{Institute of Applied Physics and Materials Engineering, University of Macau, Avenida da Universidade Taipa, Macau 999078, China}
\affiliation{Shenzhen Institute for Quantum Science and Engineering, and Department of Physics, Southern University of Science and Technology, Shenzhen 518055, China}

\author{Jian Chen}
\thanks{These authors contributed equally.}
\author{Jieming Sheng}
\thanks{These authors contributed equally.}
\author{Han Ge}
\affiliation{Department of Physics, Southern University of Science and Technology, Shenzhen 518055, China}

\author{Lianglong Huang}

\author{Cai Liu}
\affiliation{Shenzhen Institute for Quantum Science and Engineering, and Department of Physics, Southern University of Science and Technology, Shenzhen 518055, China}
\author{Zhenxing Wang}
\author{Zhongwen Ouyang}
\affiliation{Wuhan National High Magnetic Field Center and School of Physics, Huazhong University of Science and Technology, Wuhan 430074, China}

\author{Dapeng Yu}
\affiliation{Shenzhen Institute for Quantum Science and Engineering, and
	Department of Physics, Southern University of Science and Technology, Shenzhen
	518055, China}

\author{Shanmin Wang}
\affiliation{Department of Physics, Southern University of Science and Technology, Shenzhen 518055, China}

\author{Liusuo Wu}
\affiliation{Department of Physics, Southern University of Science and Technology, Shenzhen 518055, China}

\author{Hai-Feng Li}
  \email{haifengli@um.edu.mo}
\affiliation{Institute of Applied Physics and Materials Engineering, University of Macau, Avenida da Universidade Taipa, Macau 999078, China}

\author{Le Wang}
\email{wangl36@sustech.edu.cn}
\affiliation{Shenzhen Institute for Quantum Science and Engineering, and Department of Physics, Southern University of Science and Technology, Shenzhen 518055, China}

\author{Jia-Wei Mei}
\email{meijw@sustech.edu.cn}
\affiliation{Shenzhen Institute for Quantum Science and Engineering, and
   Department of Physics, Southern University of Science and Technology,
   Shenzhen 518055, China}
 \affiliation{Shenzhen Key Laboratory of Advanced Quantum Functional Materials
   and Devices, Southern University of Science and Technology, Shenzhen 518055, China}

\date{\today}

\begin{abstract}
We performed a comprehensive electron spin resonance, magnetization and heat capacity study on the field-induced magnetic phase transitions in the kagome antiferromagnet \ce{$\alpha$-Cu3Mg(OH)6Br2}. With the successful preparation of single crystals, we mapped out the magnetic phase diagrams under the $c$-axis and $ab$-plane directional magnetic fields $B$. For $B\|c$, the three-dimensional (3D) magnon Bose-Einstein condensation (BEC) is evidenced by the power law scaling of the transition temperature, $T_c\propto (B_c-B)^{2/3}$. For $B\|ab$, the transition from the canted antiferromagetic (CAFM) state to the fully polarized (FP) state is a crossover rather than phase transition, and the characteristic temperature has a significant deviation from the 3D BEC scaling. The different behaviors of the field-induced magnetic transitions for $B\|c$ and $B\|ab$ could result from the Dzyaloshinkii-Moriya (DM) interaction with the DM vector along the $c$-axis, which preserves the $c$-axis directional spin rotation symmetry and breaks the spin rotation symmetry when $B\|ab$. The 3D magnon BEC scaling for $B\|c$ is immune to the off-stoichiometric disorder in our sample \ce{$\alpha$-Cu_{3.26}Mg_{0.74}(OH)6Br2}. Our findings have the potential to shed light on the investigations of the magnetic anisotropy and disorder effects on the field-induced magnon BEC in the quantum antiferromagnet.
\end{abstract}
\maketitle

\section{Introduction}
At vanishing chemical potential and temperature, interacting bosons undergo the Bose-Einstein condensation (BEC) transition, which has been realized in ultracold dilute gases~\cite{Pethick2008} and field-induced magnons in the quantum antiferromagnet~\cite{Batiyev1984,Giamarchi2008,Zapf2014}.
Previous intense investigation of BEC in quantum magnets has been focused on dimerized materials like \ce{TlCuCl3}~\cite{Nikuni2000}, \ce{BaCuSi2O6}~\cite{Sebastian2005,Sebastian2006} and \ce{Sr3Cr2O8}~\cite{Aczel2009}, along with \ce{NiCl2}-4SC\ce{(NH2)}$_2$ holding three-dimensional-coupled $S=1$ ion-chains~\cite{Zapf2006}. Typically, quantum critical points (QCP) of transverse magnons developed into three-dimensional (3D) BEC show universal evidence $T_c(B)\propto (B_c-B)^{2/3}$~\cite{Zapf2006,Aczel2009,Radu2005,Starykh2010}. The BEC transition corresponds to the U(1) symmetry spontaneous breaking. In the bose gas system, the particle conservation guarantees the continuous U(1) symmetry whose breaking corresponds to the transition to the BEC state. Without the magnetic anisotropic interactions, the quantum antiferromagnet with the magnetic fields above the saturation field $B_{\rm c}$ has the U(1) spin rotation symmetry along the field direction, and the elementary magnetic excitation, i.e., magnon, has a gap that is controlled by the field strength. With decreasing magnetic fields, the gap vanishes at the saturation field $B_{\rm c}$, and the magnon undergoes the BEC transition with the U(1) symmetry breaking. In the real materials, the spin-orbit coupling gives rise to the magnetic anisotropic term in the quantum magnetism that breaks the U(1) symmetry and modifies the spontaneous transition to a BEC state for the magnon condensation~\cite{Zapf2014}. Since the symmetry-breaking terms are usually small enough, their effect on the magnon condensation near the saturation field is scarcely investigated.

In this work, we study the magnetic anisotopy effect on field-induced magnon condensations  in kagome antiferromaget \ce{$\alpha$-Cu3Mg(OH)6Br2}. From the previous study on the powder samples~\cite{Wei2019}, \ce{$\alpha$-Cu3Mg(OH)6Br2} has the predominant intra-layer ferromagnetic interaction characterized by the Curie-Weiss temperature $\theta =$ 34 K and the moderate inter-layer antiferromagnetic interaction corresponding to the saturated field of 2~T. Analogous to kagome ferromagnet Cu[1,3-bdc]~\cite{Chisnell2015}, \ce{$\alpha$-Cu3Mg(OH)6Br2} has the out-of-plane Dzyaloshinskii-Moriya (DM) interaction which preserves and breaks the U(1) spin rotation for $B\|c$ and $B\|ab$, respectively. As discussed theoretically in Ref.~\cite{Chernyshev2016}, the out-of-plane DM interaction gives rise to topological magnon bands when the magnetization ${\bf M}$ is parallel to the DM vector ${\bf D}$, but results into anharmonic particle-nonconserving magnon couplings when ${\bf M}\perp{\bf D}$. Since the U(1) spin rotation symmetry is crucial for the magnon BEC, the DM interacting term plays different roles in the magnon condensations near the saturation field for the $c$-axis directional and in-plane fields. Our main task in this work is to study the magnetic anisotopy effect on the magnon condensation in \ce{$\alpha$-Cu3Mg(OH)6Br2}. From the Inductively Coupled Plasma-Atomic Emission Spectrometry (ICP-AES), we have determined the ratio of Cu/Mg and found that the actual chemical formula of our sample is \ce{$\alpha$-Cu_{3.26}Mg_{0.74}(OH)6Br2}, providing an opportunity to discuss the off-stoichiometric disorder effect on the magnon condensations.

The rest of the paper is organized as the following. We present the results in the SEC.~\ref{sec:met_res}. We summarize our experimental methods for the sample growth and characterization, magnetization and heat capacity measurements and the electron spin resonance (ESR) in SEC.~\ref{subsec:methods}. From the comprehensive understanding of magnetization and ESR measurements, we write down the spin model as Eq.~(\ref{eq:1}) for \ce{$\alpha$-Cu3Mg(OH)6Br2} in SEC.~\ref{subsec:exchanges}, and the amplitudes of the interactions are justified. In SEC.~\ref{subsec:Bc}, we map out the magnetic phase diagram for $B\|c$ from the magnetization and heat capacity measurements in \ce{$\alpha$-Cu3Mg(OH)6Br2}. The 3D magnon BEC transition between the fully polarized (FP) state and the canted antiferromagetic (CAFM) state has been evidenced by the power law exponent in the transition temperature scaling $T_c\propto (B_c-B)^{2/3}$. We also discuss the magnetic off-stoichiometric disorder effect in the magnon BEC in this subsection. The field-induced phase diagram for the in-plane magnetic field presents in SEC.~\ref{subsec:Bab}. At low temperatures, from the FP state to CAFM, it is a crossover rather than a sharp phase transition. The phase boundary of the crossover characteristic temperature near the saturation field has a significant deviation from the 3D BEC scaling. SEC.~\ref{sec:concl} is the conclusion section.

\section{Results}\label{sec:met_res}
\subsection{Experimental methods}\label{subsec:methods}

The single crystals of \ce{$\alpha$-Cu3Mg(OH)6Br2} were grown by hydrothermal method. The mixture of 15~mmol \ce{CuO}, 6~mmol \ce{MgBr2}, and 10.8~mmol \ce{NH4F} was sealed in 25-mL Teflon-lined autoclave with 10 mL water. The autoclave was heated to 270 $^{\circ}$C and cooled to 140~$^{\circ}$C with a rate of 0.5 $^{\circ}$C /h. After washed with the deionized water, the blue-green
and  millimeter-scale crystals were obtained as shown in the inset of Fig.~\ref{fig:Fig1}(c). The crystal structure of \ce{$\alpha$-Cu3Mg(OH)6Br2} was determined by SuperNova Dual Wavelength diffractometer with monochromated Cu-K$\alpha$ radiation ($\lambda$ = 1.5418 \AA) at room temperature, and the program of shelXT~\cite{Sheldrick2015a} and shelXL~\cite{Sheldrick2015} were performed to solve and refine the crystal structure of \ce{$\alpha$-Cu3Mg(OH)6Br2}. The ratio of Cu/Mg was determined by ICP-AES and the chemical composition was confirmed to be $\alpha$-\ce{Cu_{3.26}Mg_{0.74}(OH)6Br2}. Beside the discussion on the off-stoichiometric disorder effect, \ce{$\alpha$-Cu3Mg(OH)6Br2} is used for the notation convenience throughout this paper.

The temperature-dependent magnetization at 0.3 T from 1.8 to 300 K and the field-dependent magnetization with magnetic field increasing and decreasing at 1.8 K were measured by Quantum Design Magnetic Property Measurement System (MPMS). To study the spin anisotropy, the temperature-dependent magnetization were accomplished along different directions ($B||b$, $B \| a^*$, and $B||c$ axes). Massive magnetization curves from 0.4 to 10 K were collected on the Hall sensor magnetometer equipped on Physical Property Measurement System (PPMS)~\cite{Cavallini2004,Candini2006}. The temperature and field dependent heat capacity measurements were performed on PPMS. The pulsed-field ESR spectra were measured in the field-increasing process among the frequency range of 54 to 172 GHz at 2 K.

\begin{table}[b]
	\footnotesize
	\caption{Crystal data and structure refinement of \ce{$\alpha$-Cu3Mg(OH)6Br2}.}
	\label{table1}
	\begin{tabular*}{0.48\textwidth}{@{\extracolsep{\fill}}ll}
		\hline
		\hline
		Formula & \ce{Cu_{3.26}Mg_{0.74}(OH)6Br2} \\
		system, space group & Trigonal, $P\bar{3}m1$ \\
		cell parameter &  $a = 6.2865 $ \AA, $c = 6.0795$ \AA\\
		Color & Blue-green  \\
		Target, wavelength  & Mo K$\alpha$, 0.71073 \AA\\
		Temperature (K) & 173 \\
		$\mu$ (mm$^{-1}$) & 17.656\\
		F (0 0 0) & 225.8\\
		$\theta$ range($\deg$) & $4.97-30.36$\\
		R($F^2$ $>$ $2\sigma(F^2)$), $wR(F^2)$, $S$ & 0.042, 0.096, 1.32 \\
		\hline
	\end{tabular*}
\end{table}

%


\subsection{Exchange interactions and spin anisotropy}\label{subsec:exchanges}
\begin{figure}[t]
	\centering
	\includegraphics[width=1\columnwidth]{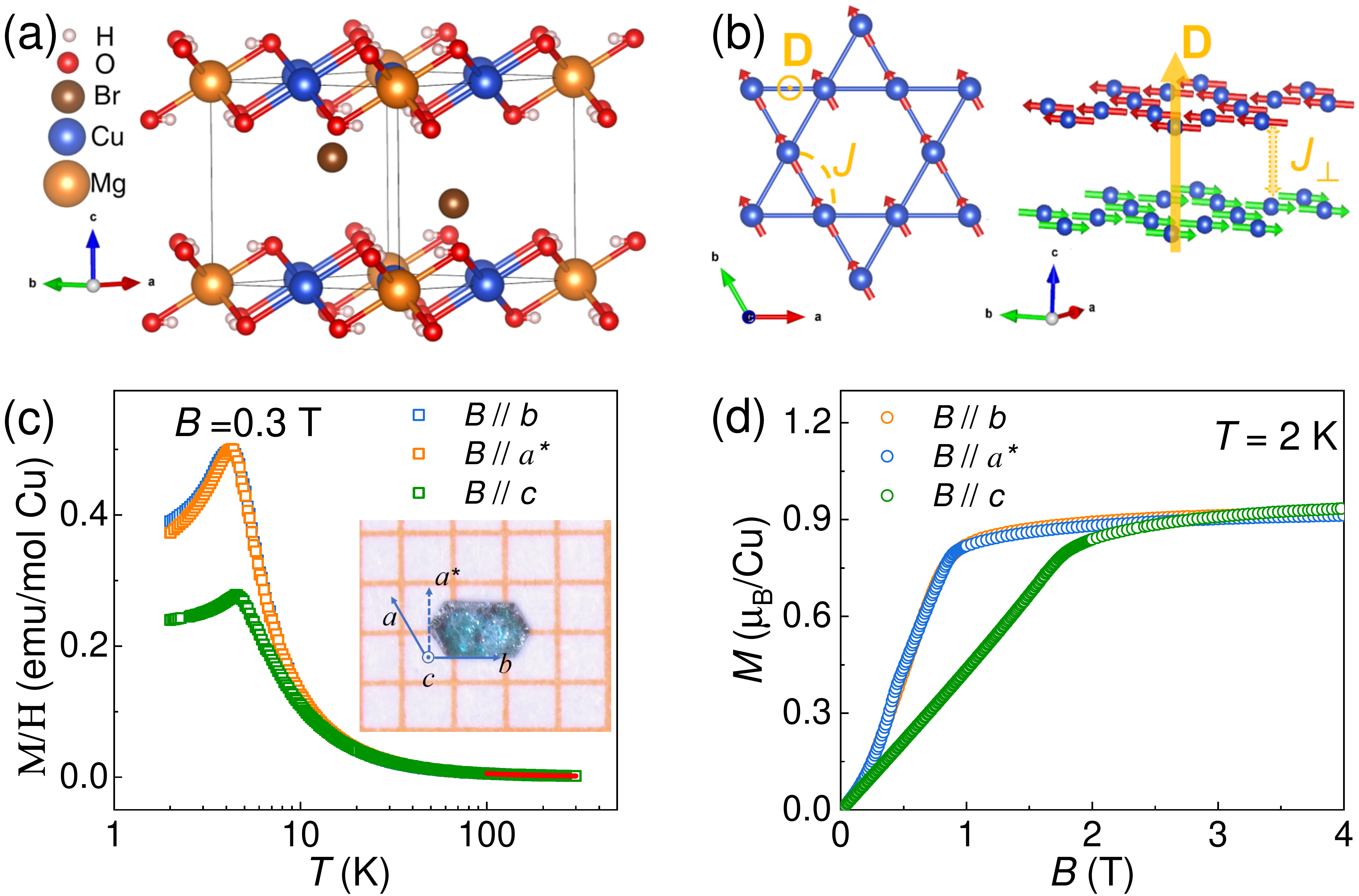}
	\caption{
    (a) The crystal structure of \ce{$\alpha$-Cu3Mg(OH)6Br2}. (b)
    \ce{$\alpha$-Cu3Mg(OH)6Br2} has the predominant intra-layer ferromagnetic
    ($J<0$)
    and the moderate inter-layer antiferromagnetic ($J_\perp>0$) interactions. The DM
    interaction has the DM vector (${\bf D}=D_z\hat{z}$) along the $c$-axis (illustrated as yellow arrow). (c)
    Temperature-dependent magnetic susceptibilities in
    \ce{$\alpha$-Cu3Mg(OH)6Br2} for fields along different directions. The red line is the
    fitting curve to magnetic susceptibilities by the Curie-Weiss law. The inset is the
    photo of the single crystal \ce{$\alpha$-Cu3Mg(OH)6Br2}. (d) Field-dependent magnetization at 2 K of \ce{$\alpha$-Cu3Mg(OH)6Br2} with fields along different directions.}
	\label{fig:Fig1}
\end{figure}

\begin{figure}[t]
	\centering
	\includegraphics[width=1\columnwidth]{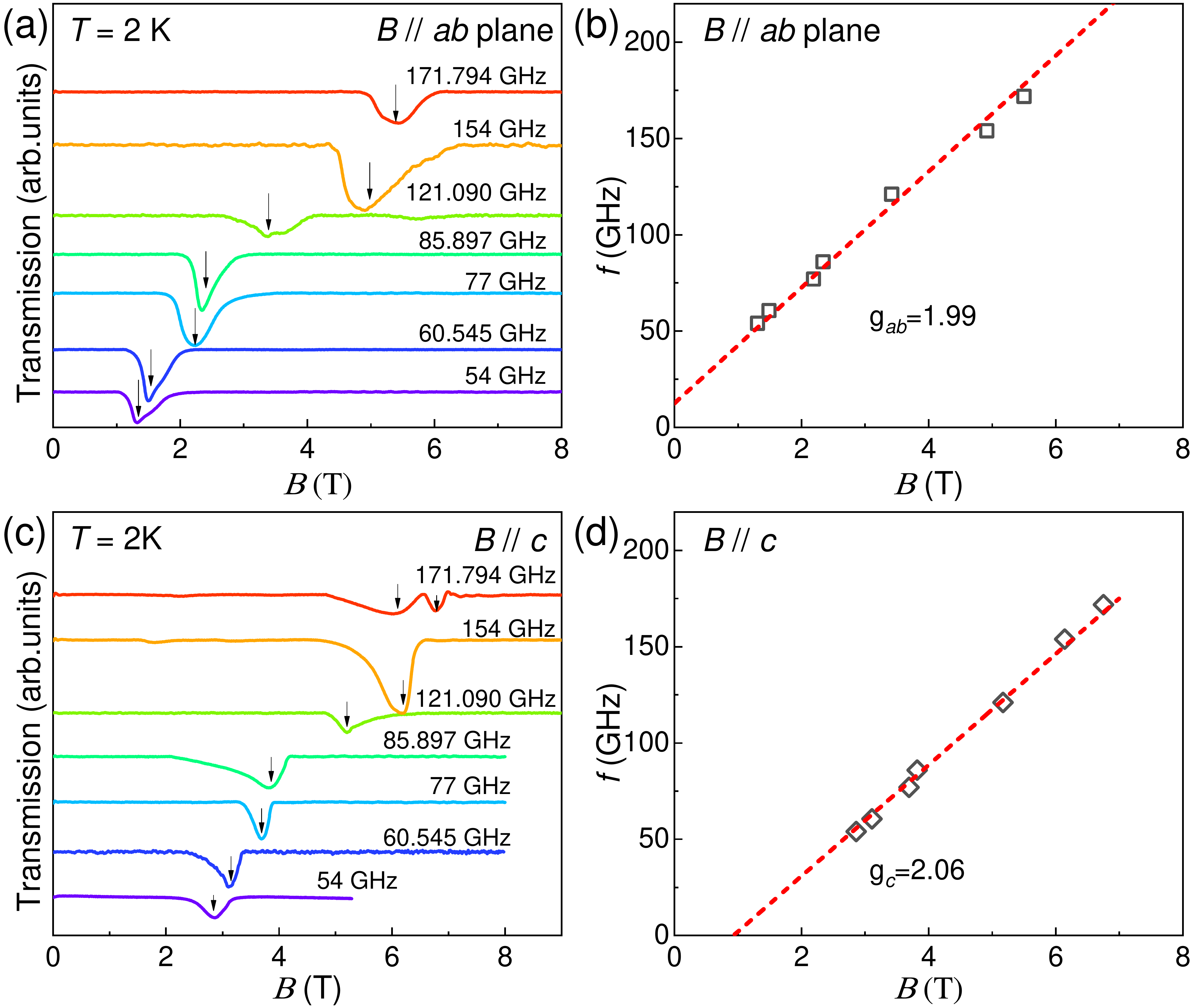}
	\caption{ESR spectra collected at 2 K in \ce{$\alpha$-Cu3Mg(OH)6Br2} for the
    in-plane field (a) and the $c$-axis directional field (c). (b) and (d)
    are the corresponding frequency-field relations obtained from the FMR
    peaks in the ESR spectra.}
	\label{fig:Fig2}
\end{figure}

As listed in Table \ref{table1}, \ce{$\alpha$-Cu3Mg(OH)6Br2} crystallizes isostructurally to haydeeite~\cite{Puphal2018}, where the Cu ions form the two-dimensional (2D) kagome lattice and the Mg ions settle into the center of hexagon. According to previous neutron scattering experiments~\cite{Wei2019}, we depict the magnetic structure of \ce{$\alpha$-Cu3Mg(OH)6Br2} as shown in Fig.\ref{fig:Fig1}(b).

Figure~\ref{fig:Fig1}(c) is the temperature dependent magnetization with fields of 0.3~T along different crystal directions. From the Curie-Weiss fitting with the $g$-factors extracted from the ESR ($g_{ab}=1.99$, and $g_c=2.06$ as shown in Fig.~\ref{fig:Fig2}), we obtain the Curie-Weiss temperatures $\theta_b=25.3$~K, $\theta_{a^*}=27.6$~K, and $\theta_c=24.4$~K, indicating the predominant ferromagnetic interactions in \ce{$\alpha$-Cu3Mg(OH)6Br2}. The Curie-Weiss temperatures are smaller than the previous reported values on the powder samples~\cite{Wei2019}, probably due to the off-stoichiometric chemical formula of our samples. There is a clear AFM ordering formed at 4.3~K, implying the AFM interlayer interactions in \ce{$\alpha$-Cu3Mg(OH)6Br2}. We see that the in-plane magnetic susceptibility is larger than the out-of-plane one at low temperatures, implying the easy-plane magnetism in \ce{$\alpha$-Cu3Mg(OH)6Br2}. Fig.~\ref{fig:Fig1}~(d) is the field dependent magnetization along different crystal directions at $T=2$~K. The easy plane and the interlayer AFM interaction can be also derived from the $M$-$B$ curves for fields along different crystal directions in Fig.~\ref{fig:Fig1}(d). Despite the zero-field spins orientation towards the $b$ axis, no obvious difference was observed between $B\|b$ and $B \|a^*$ in both the temperature-dependent and field-dependent magnetization measurements.


Figure~\ref{fig:Fig2} presents the ESR spectra collected at 2K for \ce{$\alpha$-Cu3Mg(OH)6Br2} with $B\|ab$ and $B\|c$. The dip in the spectra is the ferromagnetic resonance (FMR) mode. The slops of the field-frequency relations of the FMR modes are the $g$-factors with values of $g_{ab}=1.99$ and $g_c=2.06$. From the magnetization measurement in Fig.~\ref{fig:Fig1}, we know that the magnetism in \ce{$\alpha$-Cu3Mg(OH)6Br2} has the easy plane of the kagome plane, accounting for the positive and negative intercepts in the frequency-field-relation lines for the FMR modes with the in-plane and $c$-axis directional fields, respectively. The easy-plane anisotropic term has the order of 1~K, much smaller than the Curie-Weiss temperature with the value of $\sim$ 25~K.

From the above measurements, we can write down the spin Hamiltonian for
\ce{$\alpha$-Cu3Mg(OH)6Br2} as the following
\begin{eqnarray}
	\label{eq:1}
	H&=&J\sum_{n\langle{i,j} \rangle}\mathbf{S}_{n,i}\cdot\mathbf{S}_{n,j}+J_{\bot}\sum_{n,i}\mathbf{S}_{n,i}\cdot\mathbf{S}_{n+1,i}\nonumber\\
   &+&D\hat{z}\cdot\sum_{n\langle{i,j} \rangle}(\mathbf{S}_{n,i}\times\mathbf{S}_{n,j})-g\mu_B\textbf{\emph{B}}\cdot\sum_{n,i}\mathbf{S}_{n,i},
\end{eqnarray}
where $\mathbf{S}_{ni}$ is the spin operator on the $i$-th site in the $n$-th kagome layers, and $\langle ij \rangle$ indicates the nearest neighbor bonds in the kagome layers. The predominant FM intra-layer interaction is $J\simeq-25$~K that can be obtained from the Curie-Weiss fitting of the temperature dependent magnetization. The inter-layer AFM interaction $J_\bot$ is about 1~K. \ce{$\alpha$-Cu3Mg(OH)6Br2} has the easy-plane anisotropy in the exchange terms, however, in this work,  the DM interaction in \ce{$\alpha$-Cu3Mg(OH)6Br2} is the main magnetic anisotropy, regarding to the previous reports on the related materials~\cite{Chisnell2015,Zorko2008,Arh2020}. The DM interaction has the DM vector $\mathbf{D}=D\hat{z}$ along the $c$-axis and can be estimated around $D\simeq10$~K as we expect that the local electronic environment for the exchange path is similar to other kagome magnetic compounds.\cite{Zorko2008,Arh2020}

The DM interaction cannot generate  fluctuations in the saturated ground state, however, the same is not true for magnon excitations as discussed previously in Ref.~\cite{Chernyshev2016}. The magnon excitation spectrum is sensitive the DM interaction and depends on the relative directions of the ground state magnetization and the DM interaction vector. \ce{$\alpha$-Cu3Mg(OH)6Br2} has the DM vector along the $c$-axis. For $B\|c$ in the FP state, the magnetization is parallel to the DM vector, i.e., ${\bf M}\|{\bf D}$. The DM term provides an imaginary component to the spin-flip hoppings, resulting in the topological magnon bands as observed in neutron scattering studies on Cu[1,3-bdc]~\cite{Chisnell2015}. We stress that the DM vector along the $c$-axis preserves the U(1) spin rotation symmetry in the FP state. 
By contrast, for the in-plane field applied in the FP state, the magnetization is perpendicular to the DM vector, i.e., ${\bf M}\perp{\bf D}$, and the DM term gives rise to anharmonic interaction of magnon as it mixes the transverse magnon excitations with the longitudinal ones. The DM interaction here already breaks the U(1) spin rotation symmetry in the FP state. 

\subsection{Field-induced magnetic phase diagram for $B\|c$}\label{subsec:Bc}
\begin{figure}[t]
	\centering
	\includegraphics[width=1\columnwidth]{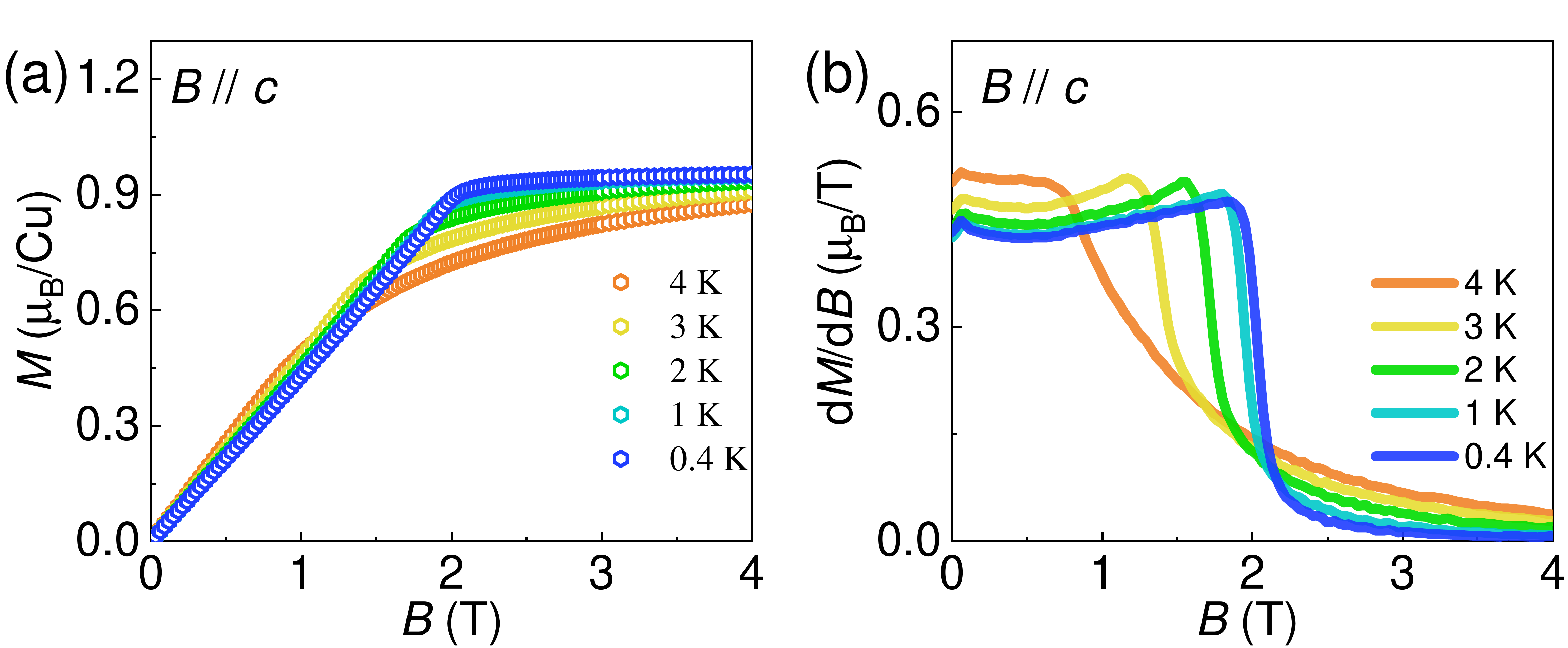}
	\caption{Field-dependent
    magnetization $M$ (a) and the corresponding differential susceptibility $dM/dB$
    curves (b), at selected temperatures for fields along the $c$-axis. }
	\label{fig:Fig3}
\end{figure}

\begin{figure}[b]
	\centering
	\includegraphics[width=0.7\columnwidth]{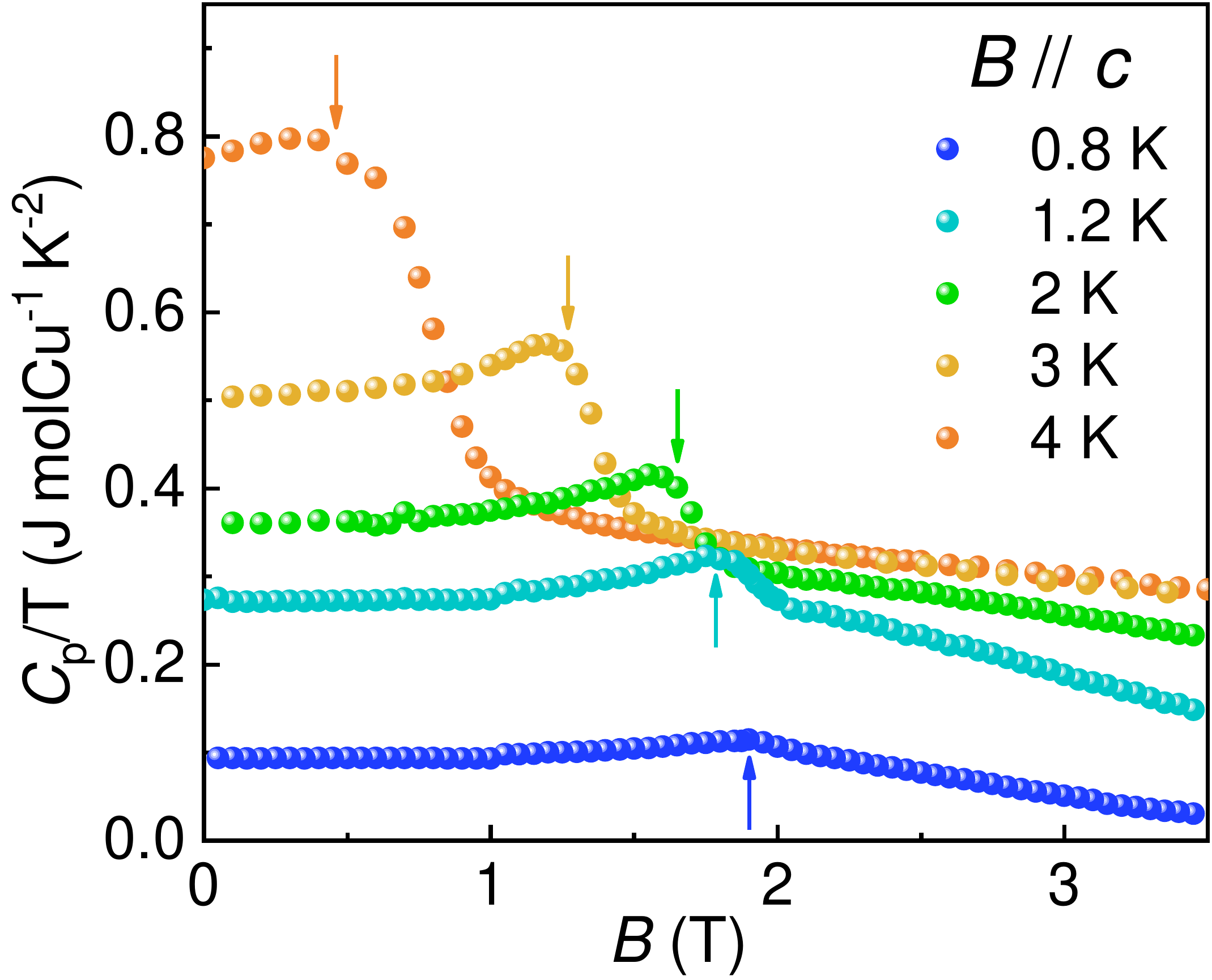}
	\caption{Field-dependent specific
    heat coefficients $C_p/T$ at selected temperatures for fields along the $c$-axis. }
	\label{fig:Fig4}
\end{figure}

\begin{figure}[t]
	\centering
	\includegraphics[width=1\columnwidth]{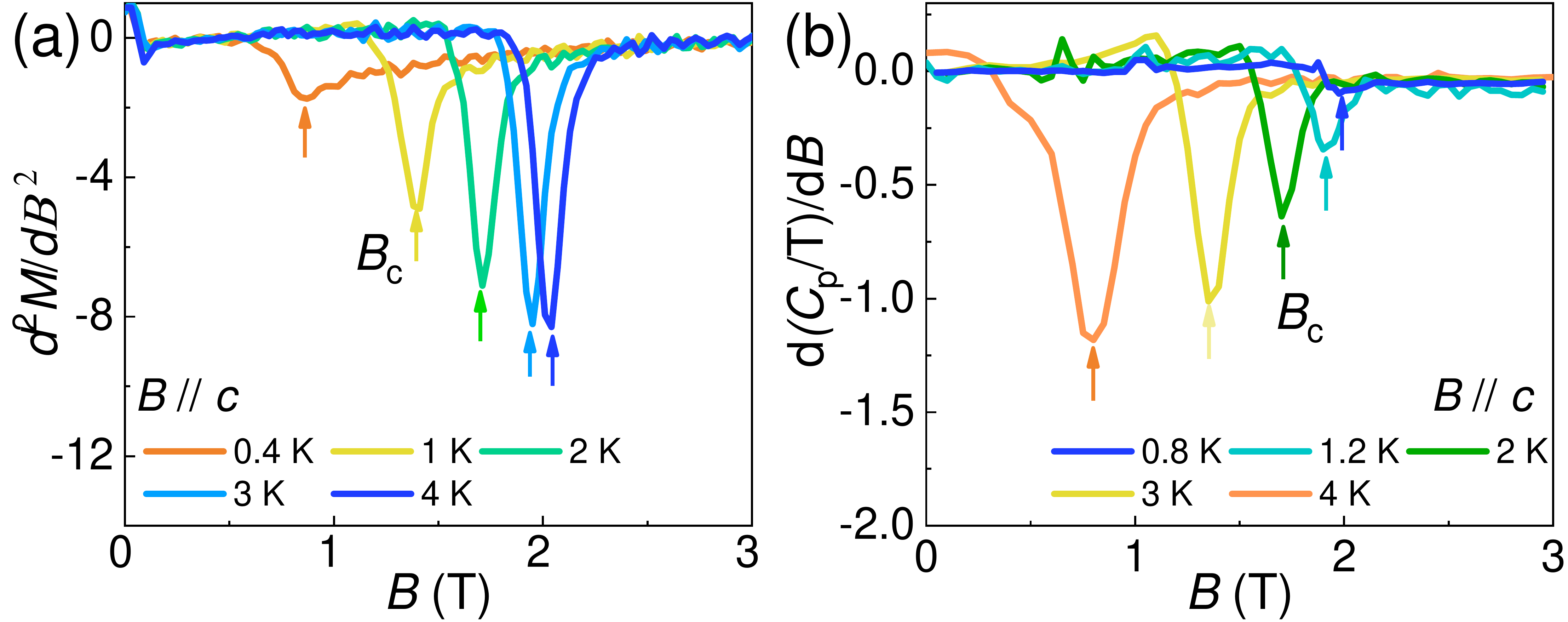}
	\caption{
    (a) Second derivative magnetization curves $\frac{d^2 M}{d
      B^2}$ at selected temperatures. (b) First derivative
    specific heat coefficients $d(C_p/T)/dB$ at selected temperatures. Arrows
    indicate the position of the saturation field $B_{\rm c}$.}
	\label{fig:Fig5}
\end{figure}

\begin{figure}[b]
	\centering
	\includegraphics[width=1\columnwidth]{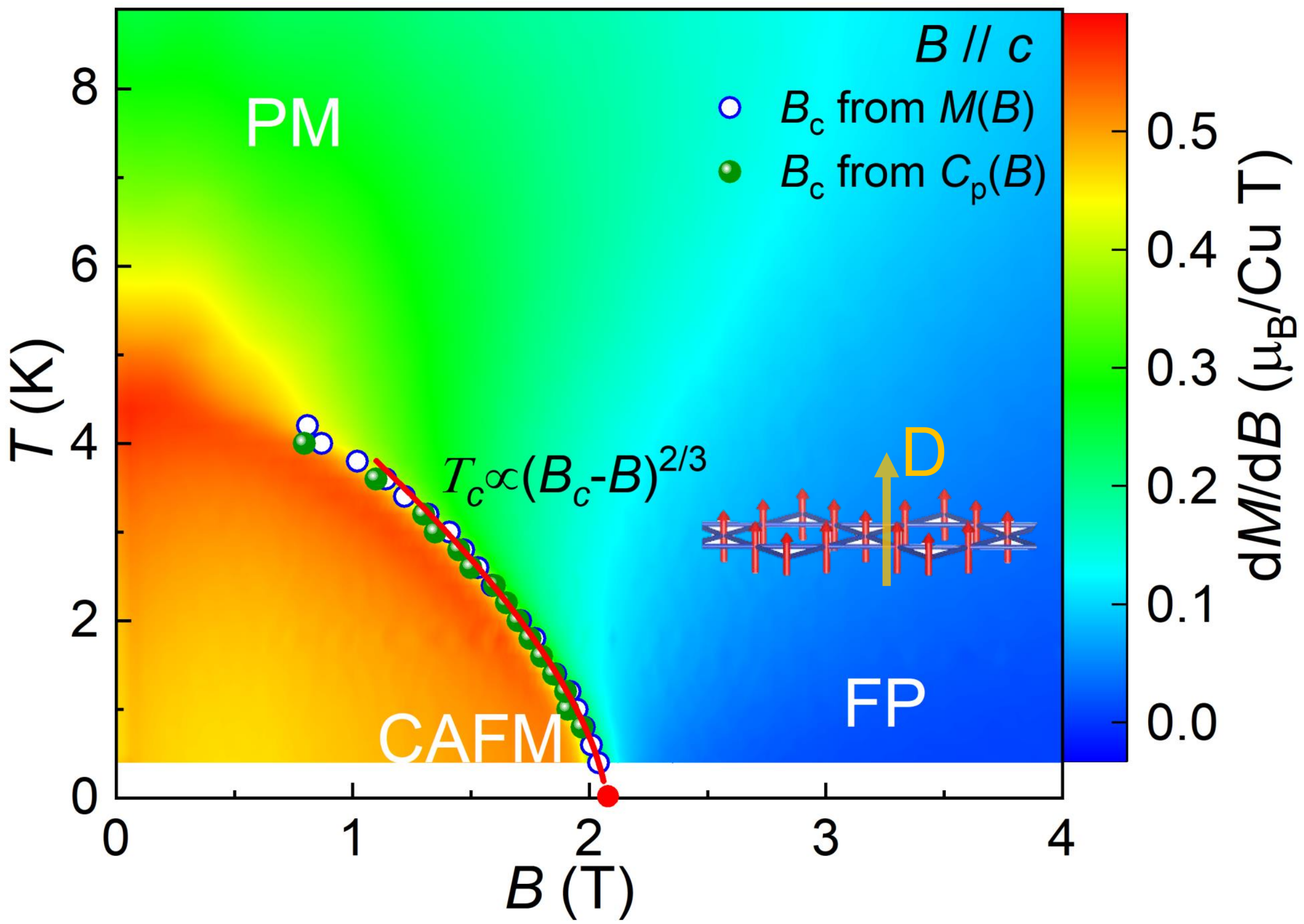}
	\caption{ The magnetic phase diagram for $B||c$. The phase boundaries determined by critical points $T_N$ and $B_c$ from $M(T)$, $C_p(T)$, $M(B)$, and $C_p(B)/T$ measurements, respectively. The red line is function of $T_c\propto (B_c-B)^{2/3}$, where $B_c$ = 2.06 T. The yellow arrow and red arrows are out-of-plane DM vector and polarized spins. 
	}
	\label{fig:Fig6}
\end{figure}

To figure out the field-induced magnetic phases for fields along the $c$-axis, we sweep the magnetic field and measure the magnetization and heat capacity for \ce{\alpha-Cu3Mg(OH)6Br2} as shown in Fig.~\ref{fig:Fig3} and Fig.~\ref{fig:Fig4}, respectively, from which the transition from the CAFM state to the FP state can be resolved, particularly from the differential results of the magnetization and heat capacity as shown in Fig.~\ref{fig:Fig5}. Figure~\ref{fig:Fig6} summarizes the field-induced phase diagram with the phase boundary between the CAFM and the FP states determined from the peaks as shown in Fig.~\ref{fig:Fig5}. The phase boundary agrees well with the 3D BEC scaling behavior $T_c\propto (B_c-B)^{2/3}$, identifying the 3D BEC quantum criticality of \ce{$\alpha$-Cu3Mg(OH)6Br2}. The 3D magnon BEC condensation behavior is consistent with the theoretical argument in SEC.~\ref{subsec:exchanges} that the
DM interaction preserves the U(1) spin rotation symmetry along the $c$-axis and the magnon BEC breaks the symmetry spontaneously.


Figure~\ref{fig:Fig3} (a) is the field-induced magnetization measurements with fields along $c$-axis at selected temperatures below the zero-field critical temperature 4.3~K. At the base temperature 0.4~K, the magnetization increases as the field increases and saturates above the saturation field $B_{\rm c}$ with the FP spins aligned along the $c$-axis. The corresponding differential susceptibility has a small bump at low fields in
Fig.~\ref{fig:Fig3}(b), indicating a crossover from the AFM state to the CAFM state. At high magnetic fields, the magnetization saturates and the differential susceptibility exhibits an abrupt jump, implying the transition from the CAFM state to the FP state. We swept the fields up and down in the magnetization measurements, and no obvious field dependent hysteresis behavior
was observed. Figure~\ref{fig:Fig4} is the field-dependent heat capacity at selected temperatures and the magnetic phase transition from the CAFM state to the FP state was also resolved.

Figures~\ref{fig:Fig5}~(a) and (b) are the the second derivative magnetization curves  and the first derivative specific heat coefficients in \ce{$\alpha$-Cu3Mg(OH)6Br2} for $B\|c$, respectively. The peak position determines the saturation field $B_{\rm c}$ for the magnetic phase transitions from the CAFM state to the FP state. Figure~\ref{fig:Fig6} is the field-induced magnetic phase diagram for $B\|c$. The intensity of the color code represents the values of the differential magnetic
susceptibility $dM/dB(B,T)$, and its boundary coincides with the saturation fields $B_{\rm c}$ extracted from the peaks as shown in Fig.~\ref{fig:Fig5}. The phase boundary was well fit by $T_c\propto (B_c-B)^{2/3}$, where $B_c = 2.06$~T, a hallmark evidence for 3D BEC of the magnon condensation.

For the FP state in \ce{$\alpha$-Cu3Mg(OH)6Br2} with fields along the $c$-axis, the DM interaction in Eq.~(\ref{eq:1}) modifies the spin-flip Hamiltonian as
\begin{eqnarray}
	\label{eq:2}
	H_{DM}^{||}&=& \frac{\widetilde{J}}{2}\sum_{n\langle{i,j}
                 \rangle}(S_{n,i}^+S_{n,j}^-e^{i\theta}+S_{n,i}^-S_{n,j}^+e^{-i\theta})\nonumber\\
  &+&\frac{J_\perp}{2}\sum_{n,i}(S_{n,i}^+S_{n+1,i}^-+S_{n,i}^-S_{n+1,i}^+),
\end{eqnarray}
with $\widetilde{J}=-\sqrt{J^2+D^2}$, and $\theta=\arctan(D/|J|)$. The DM interaction gives an imaginary phase for the spin-flip hopping with the kagome plane and still preserves the U(1) spin rotation symmetry along the $c$-axis. The magnon excitation for the spin-flip Hamiltonian in Eq.~(\ref{eq:2}) gives rise to exotic topological magnon bands~\cite{Chisnell2015,Chernyshev2016} which deserve future neutron scattering studies of \ce{$\alpha$-Cu3Mg(OH)6Br2}. The U(1) spin rotation symmetry breaking results in the 3D magnon BEC transition near the saturation field in \ce{$\alpha$-Cu3Mg(OH)6Br2}.

Having established the 3D magnon BEC in \ce{$\alpha$-Cu3Mg(OH)6Br2}, we can now use the compound to check the disorder effect in BEC. Since the chemical doping could bring about dirty bosons, the BEC phase boundary may shift and critical exponent can be further changed~\cite{Zheludev2013,yao2014,wulf2013}. The
dimensionality of the magnon BEC in \ce{$\alpha$-Cu3Mg(OH)6Br2} is $d=3$. The dynamical exponent for the magnon excitation is $z=2$ according to the Eq.~(\ref{eq:2}), and then the critical scaling exponent for the correlation length $\nu=1/2$ regarding to $\nu z=1$ due to the symmetry property~\cite{Sachdev1994}. Therefore, the BEC quantum criticality here satisfies the Harris criterion $d\nu<2$, implying the disorder is irrelevant for the quantum phase transition~\cite{Harris1974}. From ICP-AES, we find that the actual chemical formula of our sample is \ce{$\alpha$-Cu_{3.26}Mg_{0.74}(OH)6Br2}, however, the
off-stoichiometric magnetic disorder does not change the critical behavior of the 3D magnon BEC as shown in Fig.~\ref{fig:Fig6}.


\subsection{Field-induced phase diagram for $B\|ab$}\label{subsec:Bab}

\begin{figure}[b]
	\centering
	\includegraphics[width=1\columnwidth]{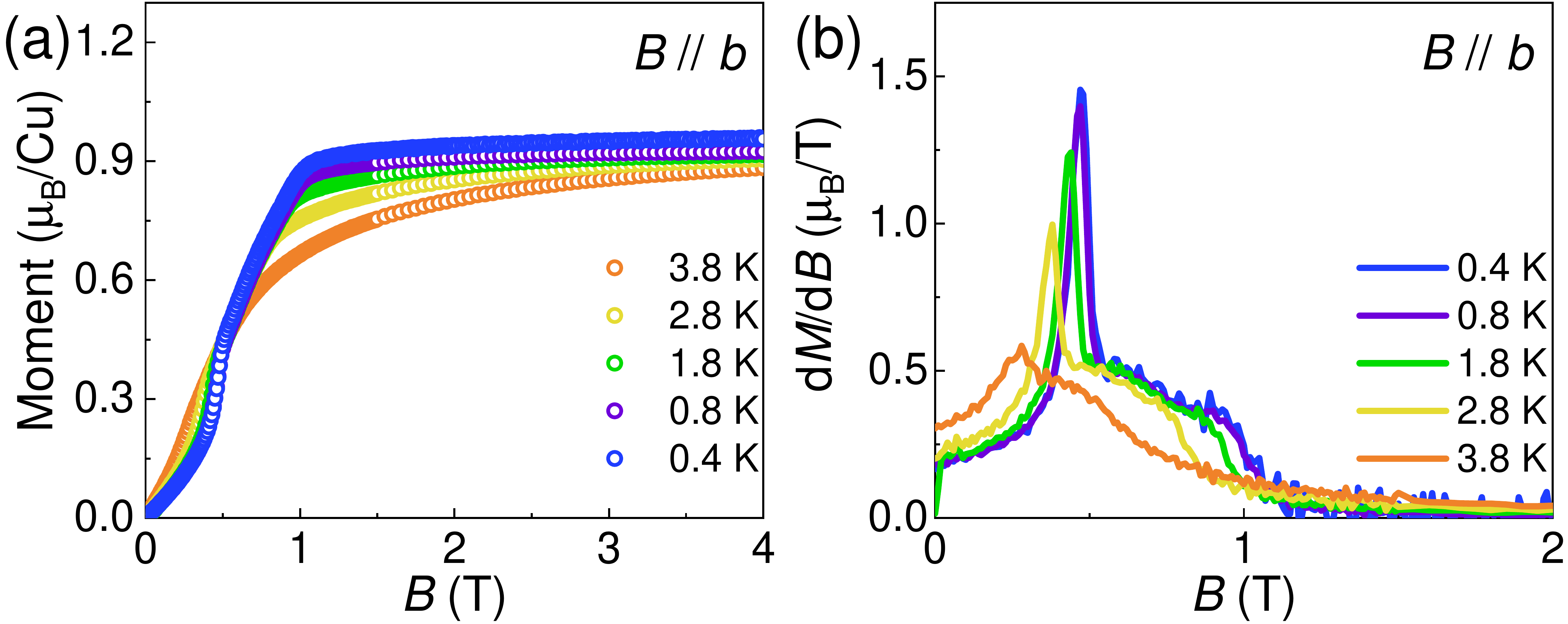}
	\caption{Field-dependent magnetization $M$ and corresponding differential
    susceptibility $dM/dB$ curves in (a) and (b), respectively, at selected
    temperatures for the in-plane fields.}
	\label{fig:Fig7}
\end{figure}

\begin{figure}[t]
	\centering
	\includegraphics[width=1\columnwidth]{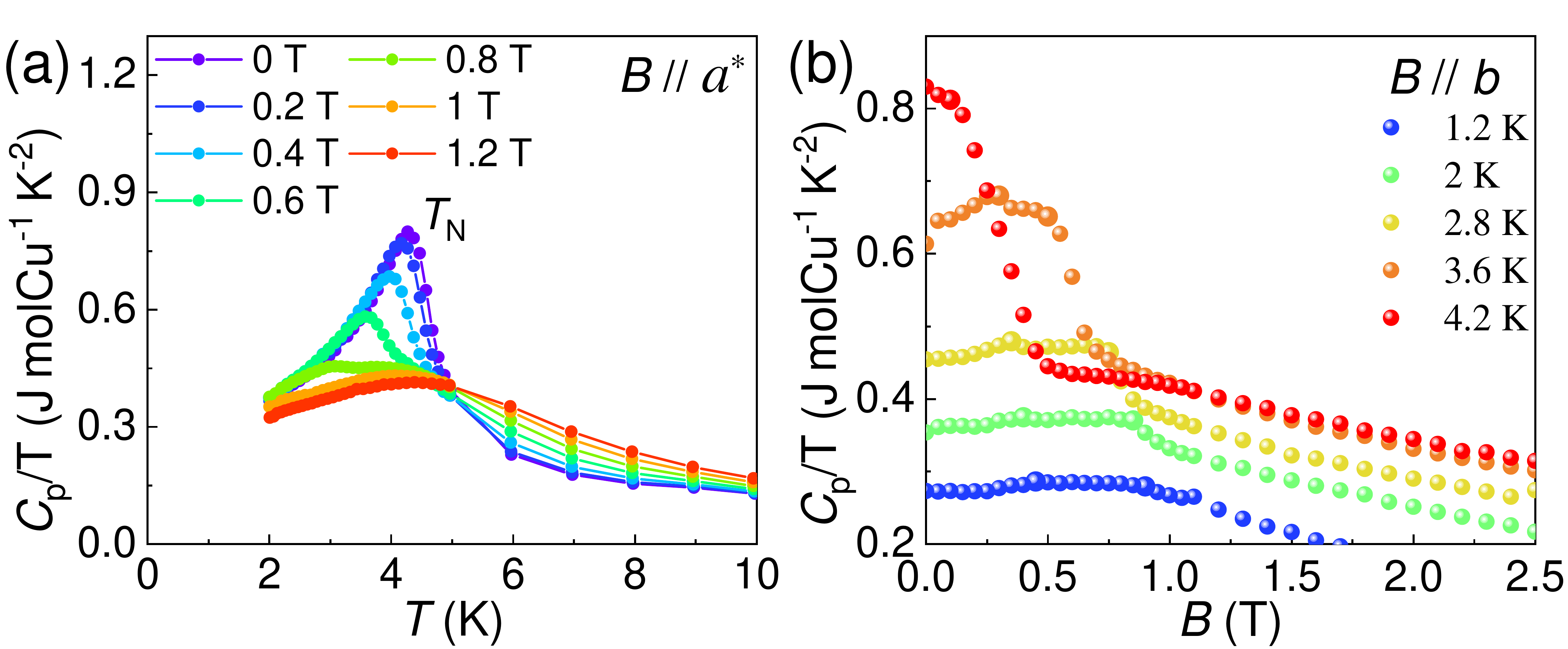}
	\caption{(a) Temperature-dependent  heat capacity for $B \| a^*$ with selected
    fields. (b) Field-dependent specific heat for $B\|b$ at selected temperatures.}
	\label{fig:Fig8}
\end{figure}

\begin{figure}[b]
	\centering
	\includegraphics[width=0.7\columnwidth]{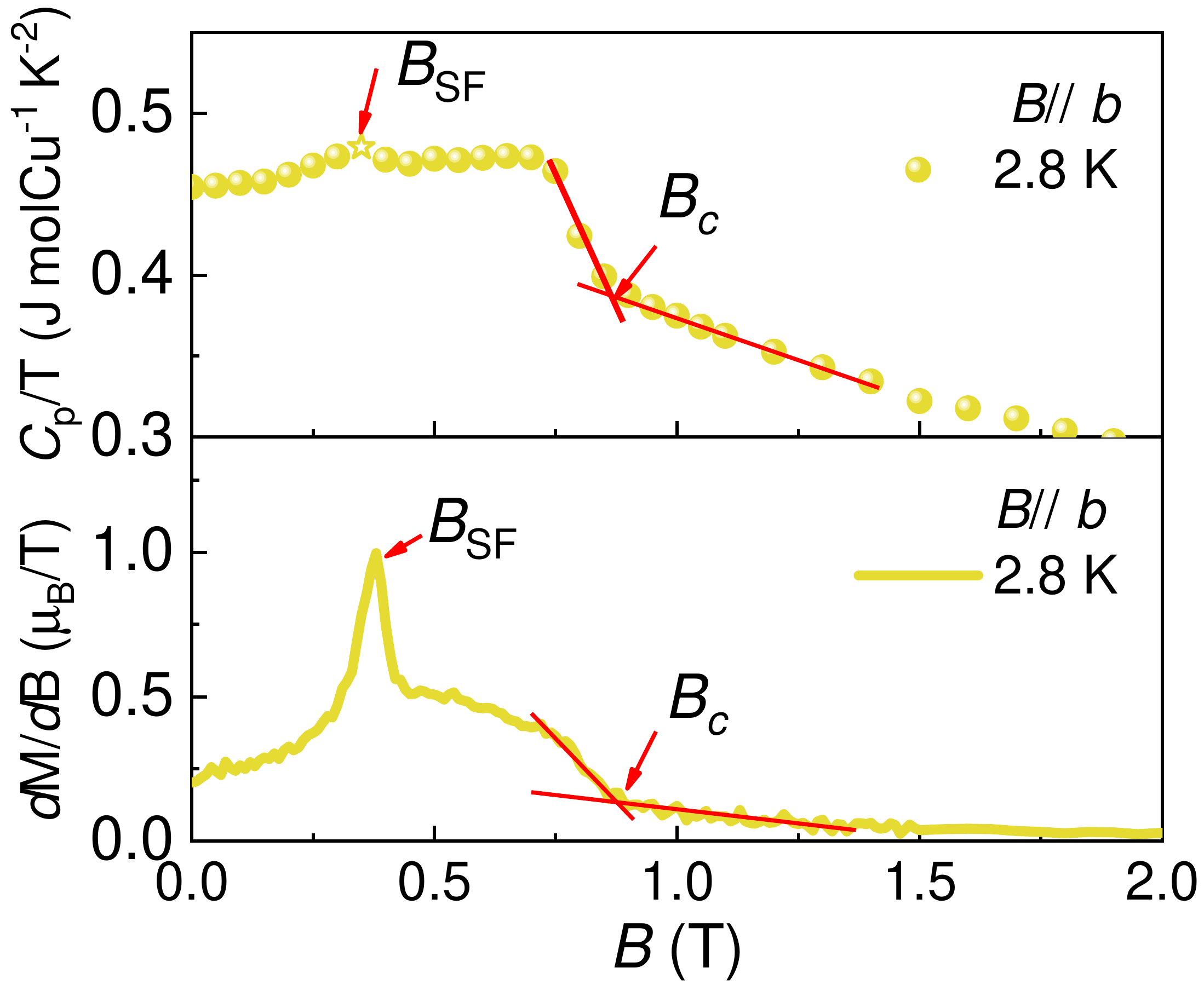}
	\caption{Characteristic critical field $B_c$ for the crossover from the CAFM
    state to the FP state.
	}
	\label{fig:Fig9}
\end{figure}

\begin{figure}[t]
	\centering
	\includegraphics[width=1\columnwidth]{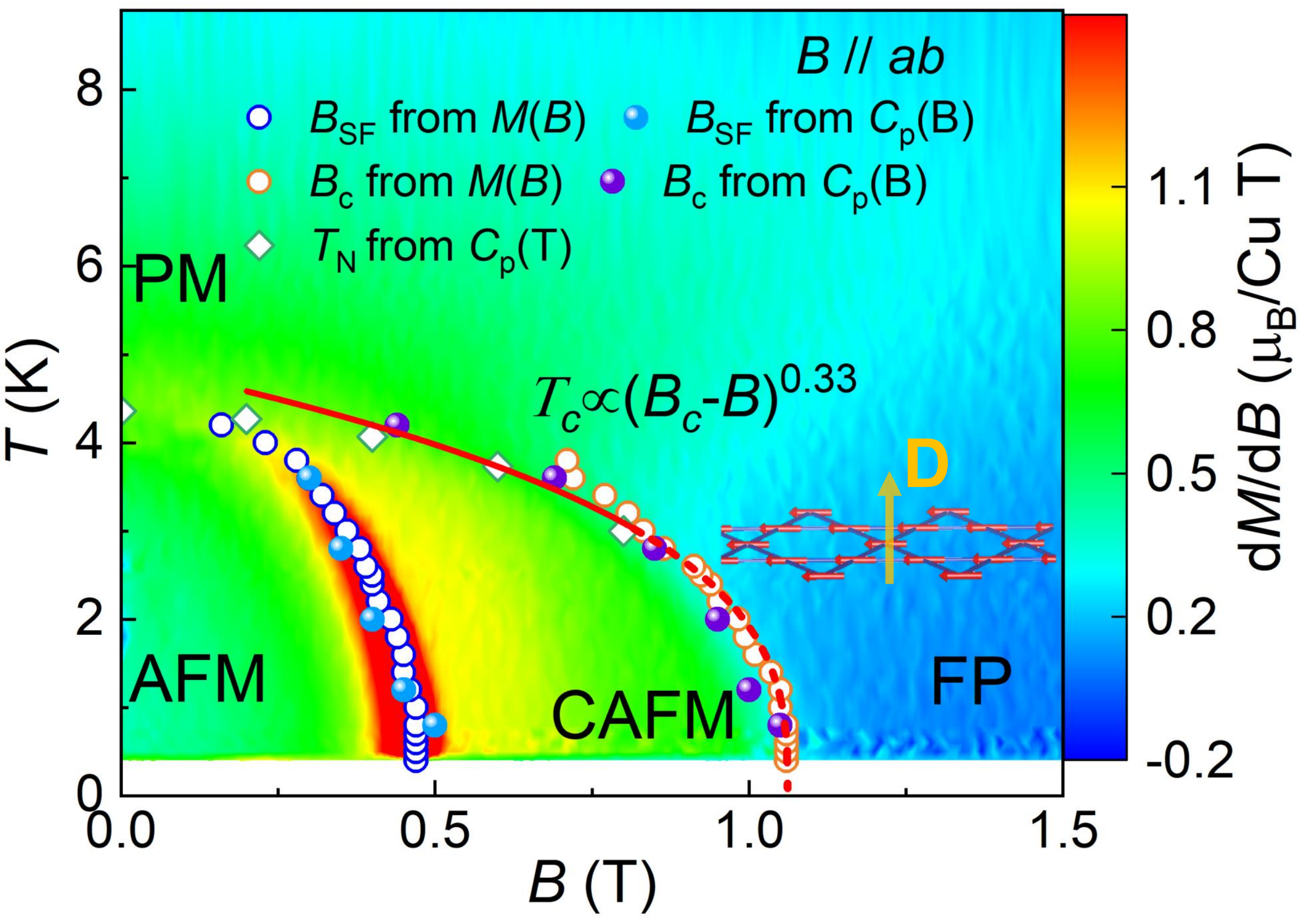}
	\caption{The in-plane field-induced magnetic phase diagram.
		The red solid and dashed line is fitted by $T_c\propto (B_c-B)^{\delta}$, with $ \delta=0.33 $ and $ B_c=1.06 $ T. The yellow arrow and red arrows are out-of-plane DM vector and polarized spins. }
	\label{fig:Fig10}
\end{figure}

To figure out the field-induced magnetic phases for the in-plane fields, we performed the magnetization and heat capacity measurements for \ce{\alpha-Cu3Mg(OH)6Br2} with fields parallel to $b$-axis, as shown in Fig.~\ref{fig:Fig7} and Fig.~\ref{fig:Fig8}, respectively. Figure~\ref{fig:Fig7} (a) is the in-plane field-dependent magnetization curves at selected temperatures below $T_N$ of 4.3~K. At $T$ = 0.4~K, an obvious spin-flop transition emerges at $B_{\rm SF} \sim 0.5$ T. With further increasing the fields, the magnetization becomes saturated at $B_{\rm c} \sim$ 1 T. The corresponding differential susceptibility, as shown in Fig.~\ref{fig:Fig7} (b), has a peak at $B_{\rm SF}$ and a smooth shoulder at $B_{\rm c}$, implying a sharp SF transition from the AFM state to the CAFM state and a crossover from the CAFM state to the FP state, respectively.



Figure~\ref{fig:Fig8}(a) is the temperature-dependent heat capacity for $B\|a^*$ under low fields. A peak-like anomaly at about 4 K is suppressed to low temperatures by fields, corresponding to the AFM phase transition. The peak is hardly resolved above 0.8 T and evolves into another broad peak that we attribute as a crossover. Fig.~\ref{fig:Fig8} (b) is the field-dependent specific heat for $B\|b$ at selected temperatures, in which there is a sudden drop and a smooth crossover at high and low temperatures, respectively. Combined the result of magnetization, the characteristic critical field $B_{\rm c}$ is determined by two crossed lines as shown in Fig.~\ref{fig:Fig9}.

Figure~\ref{fig:Fig10} summarizes the in-plane field-induced magnetic phase diagram  of \ce{$\alpha$-Cu3Mg(OH)6Br2}. The color intensity represents the value of the differential magnetic susceptibility $dM/dB(B,T)$ as a function of $B$ and $T$. At high temperatures and low fields, it is a phase transition from the CAFM state to the PM/FP state as represented by the solid line. At low temperatures and high fields, such a transition turns out to be a crossover as represented by the dashed line. The solid and dashed line exhibits a scaling behavior $T_c\propto (B_c-B)^{\delta}$ with $\delta=0.33$, $B_c=1.06$, and a significant deviation from the 3D BEC scaling as shown in Fig.~\ref{fig:Fig6}.

For the in-plane FP state in \ce{$\alpha$-Cu3Mg(OH)6Br2}, the DM interaction vector is perpendicular to the magnetization and gives rise to the term~\cite{Chernyshev2016}
\begin{eqnarray}
  \label{eq:3}
  H_{\rm DM}^{\perp} =\frac{D}{2}\sum_{n\langle{i,j} \rangle}[(S_{n,i}^++S_{n,i}^-)S_{n,j}^z-S_{n,i}^z(S_{n,j}^++S_{n,j}^-)],
\end{eqnarray}
which is a cubic term if we rewrite the spin-flip operator in terms of the Matsubara-Matsuda magnon representation~\cite{Matsubara1956}. $H_{DM}^{\perp}$ is the symmetry-breaking term of uniaxial-symmetry and its effect on the magnon spectrum has been carefully studied in Ref.~\cite{Chernyshev2016}. Such a symmetry breaking term accounts for the crossover between the CAFM state and the FP state in \ce{$\alpha$-Cu3Mg(OH)6Br2} for the in-plane fields, and we urge further theoretical investigations on this point.

\section{Conclusions}\label{sec:concl}
we have systematically studied field-induced phase transition in the kagome antiferromaget \ce{$\alpha$-Cu3Mg(OH)6Br2} and depicted out the $c$-axis and $ab$-plane magnetic phase diagrams based on the thermodynamic properties. With $B\|c$-axis, a three-dimensional (3D) magnon Bose-Einstein condensation (BEC) occurs at the saturation field $B_{\rm c}$ between fully polarized state and the canted antiferromagnetic state and is demonstrated by the power law scaling of the transition temperature, $T_c\propto (B_c-B)^{2/3}$. With $B\|ab$-plane, it is a crossover rather than a phase transition, deviating from 3D BEC scaling. The Dzyaloshinkii-Moriya interaction with the DM vector along the $c$-axis in \ce{$\alpha$-Cu3Mg(OH)6Br2} acts as  as `` on/off "  effect on the particle-conservation term for $M \| c$ and $M \perp c$, respectively, and accounts for the different behaviors of the field-induced magnetic phase transitions with the $c$-axis and in-plane fields.

\acknowledgments
The work at SUSTech was partially supported by the program for Guangdong
Introducing Innovative and Entrepreneurial Teams (No. 2017ZT07C062), Shenzhen
Key Laboratory of Advanced Quantum Functional Materials and Devices (No.
ZDSYS20190902092905285), Guangdong Basic and Applied Basic Research Foundation
(No. 2020B1515120100), and China Postdoctoral Science Foundation (2020M682780).
Z.W.Ouyang is grateful for the funding of U20A2073.
The work at University of Macau was supported by the Science and Technology Development Fund, Macao SAR (File No. 0051/2019/AFJ), Guangdong Basic and Applied Basic Research Foundation (Guangdong--Dongguan Joint Fund No. 2020B1515120025), and Guangdong--Hong Kong--Macao Joint Laboratory for Neutron Scattering Science and Technology (Grant No. 2019B121205003).

\bibliography{./refs}

\end{document}